\documentclass[10pt,twocolumn,twoside]{IEEEtran}
\IEEEoverridecommandlockouts

\usepackage{cite}
\usepackage{amsmath,amssymb,amsfonts}
\usepackage{algorithmic}
\usepackage{graphicx}
\usepackage{textcomp}
\usepackage{balance}
\usepackage{xcolor, soul}
\usepackage[utf8]{inputenc}

\usepackage{float}
\usepackage{amsthm}
\usepackage[english]{babel}
\usepackage{verbatim}
\usepackage{comment}
\newtheorem{theorem}{Theorem}

\newtheorem{assumption}{Assumption}
\newtheorem{proposition}{Proposition}
\newtheorem{definition}{Definition}

\usepackage[ruled]{algorithm2e}
\usepackage{ulem}
\usepackage{pstricks}
\usepackage{xurl}
\usepackage{bm}
\usepackage{bbm}
\usepackage{mathtools}

\DeclareFontFamily{OT1}{pzc}{}
\DeclareFontShape{OT1}{pzc}{m}{it}{ <-> s*[1.15] pzcmi7t }{}
\DeclareMathAlphabet{\mathpzc}{OT1}{pzc}{m}{it}

\allowdisplaybreaks

\usepackage{titlesec}
\titlespacing*{\section}{0pt}{*2}{*0.8}

\newcommand\bbE{\ensuremath{\mathbb{E}}} 

\newcommand{\mat}[1]{\boldsymbol{#1}}
\renewcommand{\vec}[1]{\boldsymbol{\mathrm{#1}}}
\newcommand{\vecalt}[1]{\boldsymbol{#1}}

\newcommand{\normof}[1]{\|#1\|}

\newcommand{\abs}[1]{\vert #1 \vert}

\newcommand{\RR}{\mathbb{R}}

\newcommand{\LL}{\mathcal{L}}

\providecommand{\mD}{\ensuremath{\mat{D}}}

\providecommand{\vpi}{\ensuremath{\vecalt{\pi}}}

\providecommand{\vlambda}{\ensuremath{\vecalt{\lambda}}}
\providecommand{\Ttrain}{\ensuremath{T_\text{train}}}
\providecommand{\Thour}{\ensuremath{T_\text{hour}}}
\providecommand{\Tday}{\ensuremath{T_{\text{day}}}}
\providecommand{\vbelief}{\ensuremath{\hat{\vlambda}}}
\providecommand{\belief}{\ensuremath{\hat{\lambda}}}
\providecommand{\nd}{\ensuremath{q}}
\providecommand{\reg}{\text{REG}}

\providecommand{\vLL}{\ensuremath{\vec{\LL}}}

\def\BibTeX{{\rm B\kern-.05em{\sc i\kern-.025em b}\kern-.08em
    T\kern-.1667em\lower.7ex\hbox{E}\kern-.125emX}}
\usepackage{lipsum}

\title{\Large \bf A Hybrid Mean Field Framework for Aggregators Participating in Wholesale Electricity Markets}

\author{Jun~He 
        and~Andrew~L.~Liu
\thanks{Jun He is with Edwardson School of Industrial Engineering, Purdue University, West Lafayette, IN, USA, email: he184@purdue.edu.}
\thanks{Andrew L. Liu is with Edwardson School of Industrial Engineering, Purdue University, West Lafayette, IN, USA, email: andrewliu@purdue.edu.}}

\begin{document}
\begingroup
\allowdisplaybreaks

\maketitle

\begin{abstract}
The rapid growth of distributed energy resources (DERs) is reshaping wholesale electricity markets, where aggregators coordinate large populations of prosumers (aka DER owners) to participate at scale. A key challenge is that aggregators are typically modeled as price takers, yet their collective charging and discharging decisions can meaningfully influence market prices through system balance and network constraints.

Most existing approaches either optimize a single aggregator under exogenous prices or analyze a small number of strategic agents. Neither captures the scale, structure, or price feedback inherent in real DER integration. We address this gap with a hybrid mean field framework that models a very large prosumer population through its infinite agent limit, coordinated by a finite number of aggregators participating in wholesale markets. Each aggregator maximizes the collective payoff of its prosumers, a cooperative structure well represented by mean field control. Across aggregators, interactions are non-cooperative; however, as a starting point, we assume aggregators remain price takers rather than engaging in strategic behavior. Market prices remain endogenous through aggregate prosumer actions, leading to a mean field equilibrium. We establish conditions under which this equilibrium exists and is unique.

To handle uncertainty in demand, renewable generation, and market prices, we design a two-phase reinforcement learning algorithm that enables aggregators to learn optimal control policies. We demonstrate the framework on the power system of Oahu. Results show reduced price volatility, flatter demand profiles with improved storage utilization, and lower costs for both consumers and prosumers, providing an algorithmic foundation for integrating decentralized DERs through aggregators into wholesale electricity markets.
\end{abstract}


\section{Introduction}
\label{sec:intro}

Distributed energy resources (DERs), such as rooftop solar and energy storage, are growing rapidly in adoption, enabling end-users to act as prosumers who both produce and consume electricity. Through bi-directional power flows, these prosumers are shifting the grid from a centralized to a more decentralized structure.

To accommodate this transition, FERC Order 2222 enables DER participation in wholesale markets. Because individual DERs are typically too small to meet market requirements, aggregators, also known as virtual power plants (VPPs), serve as intermediaries. Companies such as OhmConnect and Tesla already aggregate large numbers of households or devices into wholesale market resources \cite{OhmConnect,Tesla}. Integrating many heterogeneous, small-scale DERs at this scale raises new challenges in understanding their collective impact on market prices, volatility, and system reliability.

Most existing work studies how a single aggregator optimally manages a DER portfolio under exogenous wholesale electricity prices, known as locational marginal prices (LMPs) \cite{contreras2017participation,Birge,RobustBidding}. While useful for operational insights, these models overlook that under large-scale DER participation, aggregators’ collective actions affect price formation. At scale, this creates a closed-loop feedback structure between aggregator actions and LMPs that must be captured to design algorithms in shaping market outcomes under high DER penetration.

To address this need, we develop a prescriptive, learning-based framework grounded in mean-field games (MFGs), where each aggregator optimizes its own portfolio decisions in response to dynamic market outcomes. The goal is not to mimic or study existing market behavior, but to engineer a scalable and decentralized system in which aggregators learn to optimize their DER portfolios under endogenously determined prices. Reinforcement learning (RL) enables each aggregator to adapt its strategy over time. We establish conditions for the existence and uniqueness of a mean-field equilibrium (MFE), which characterizes the steady-state outcome of this learning process. By explicitly modeling the feedback between aggregator actions and LMPs, the framework provides a foundation for AI-enabled, automated participation of DER aggregators in wholesale markets.

While MFGs capture interactions across aggregators, each aggregator must still manage a large population of DERs internally. To ensure scalability, we adopt a mean-field control (MFC) formulation in which the aggregator maximizes the collective payoff of its prosumers. Energy storage introduces intertemporal coupling, and variability in weather, load, and market prices creates significant uncertainty. Classical optimization methods rely on centralized coordination or accurate probabilistic models, whereas the MFC approach learns optimal policies directly from interaction with the environment. By representing prosumers through a representative agent in the infinite-population limit, the approach substantially reduces computational complexity while enabling adaptive, forward-looking control.

The contributions of this paper are twofold. First, we develop a hybrid MFC–MFE framework integrated with RL that enables decentralized DER aggregators to participate in wholesale electricity markets under uncertainty. Second, we design a two-phase RL algorithm for the hybrid MFC–MFE model and demonstrate its effectiveness on the Oahu power system, showing reduced price volatility, stabilized aggregator behavior, and lower customer costs.

The remainder of the paper is organized as follows. Section~\ref{sec:lit} reviews related literature. Section~\ref{sec:Wholesale} presents the wholesale market model. Section~\ref{sec:Agg} formulates the aggregator's problem under the hybrid MFC–MFE framework. Section~\ref{sec:alg} introduces the two-phase RL algorithm. Section~\ref{sec:numerical} reports numerical experiments, and Section~\ref{sec:conclusion} concludes.

\section{Literature Review} \label{sec:lit}

A growing body of work examines DER participation in wholesale electricity markets using multi-agent decision making and game-theoretic methods. Early studies generally treat each participant as an individual decision maker and rely on optimization- or learning-based formulations. For instance, \cite{IRIA20191361} proposes a two-level optimization model for a single aggregator, and \cite{ye-ddpg-single-bidding} applies deep RL for individual bidding. Related multi-agent approaches include \cite{fallahi2019linear}, which uses linear programming for demand response, and \cite{shafie2014stochastic}, which develops a multilayer agent-based model. Institutional designs enabling aggregator participation are analyzed in \cite{chen2024wholesale}. Learning-based approaches, particularly multi-agent reinforcement learning (MARL), have also been used in energy systems, though mostly in peer-to-peer or small-scale contexts \cite{liu2021multi}.

Although MARL is related to game-theoretic notions such as Nash equilibria, convergence guarantees are rare in stochastic games, and equilibrium computation becomes intractable as the number of agents increases \cite{zhang2021multi}. These challenges motivate mean-field approximations, where interactions in a large population are modeled through a representative agent interacting with the population distribution. The resulting MFE framework greatly improves tractability while approximating behavior in large but finite systems.

Recent work explores mean-field ideas under both known and unknown system dynamics. With known dynamics, \cite{feng2025decentralized} solves MFGs via dynamic programming. Under unknown dynamics, \cite{guo2019learning} and \cite{xie21g-lwp} develop RL-based MFG algorithms for competitive agents, although these typically require either full transition knowledge or direct access to the mean field. The sandbox method in \cite{sandbox} avoids a mean-field oracle by learning from a single sample path, but it operates in generic single-population environments without market-clearing feedback. Meanwhile, \cite{mondal2023mean} applies MFC control to approximate cooperative MARL settings, but does not model endogenous price formation or interactions across multiple aggregators.

In contrast, our work develops a mean-field framework tailored to DER aggregators participating in a wholesale electricity market. Each aggregator solves a cooperative MFC problem over a large DER population, aggregators interact indirectly through market clearing, and learning proceeds without access to the true mean field or uncertainty distributions, relying only on observed LMPs. This structure captures the closed-loop price effects absent in prior MFG/MFC RL approaches while preserving scalability.

\section{Wholesale Market Model and Locational Marginal Prices}\label{sec:Wholesale}
This section presents a standard wholesale electricity market model along with the formulation of LMPs. We consider a wholesale electricity market operating over a transmission network with $N$ buses, $L$ transmission lines, and $G$ bulk generators. Each bus $n \in \mathcal{N}\coloneq\{1, \ldots, N\}$ serves a group of $M_n$ agents, comprising $M_n^p$ prosumers and $M_n^c$ pure consumers without any DERs. Each bulk generator $g \in \mathcal{G}\coloneq\{1, \ldots, G\}$ is characterized by a cost function $C_g(\cdot)$ that captures the cost of electricity production. Let $G_n \subseteq \mathcal{G}$ denote the set of generators connected to bus $n$. We require the sets to be disjoint and collectively exhaustive; that is,  $\cup_{n=1}^N G_n = \mathcal{G}$ and $G_n \cap G_{n'} = \emptyset$ for any $n \ne n'$.


On the demand side, let \( d^n_{it} \) denote the net demand of prosumer \( i \) at bus \( n \), defined as total energy consumption minus local solar generation. If \( d^n_{it} > 0 \), the prosumer is a net consumer; if \( d^n_{it} < 0 \), the prosumer exports energy to the grid. This bidirectional interaction allows prosumers to participate flexibly in the energy market. Similarly, we use \( d^n_{jt} \geq 0 \) to denote the demand of consumer \( j \) at bus \( n \). The total demand at bus \( n \) and time \( t \) is:
\begin{equation}
    D^n_t = \sum_{i=1}^{M^p_n} d^n_{it} + \sum_{j=1}^{M^c_n} d^n_{jt}. \label{eq:Demand_n}
\end{equation}
To avoid technical complications from infeasible supply-demand imbalances, we assume that total net demand is non-negative at each timestep $t$: $\sum_{n=1}^{N} D^n_t \geq 0$.

The wholesale market operates over discrete time intervals, indexed by \( t \in \{1, 2, \ldots\} \), where each timestep corresponds to a fixed duration, such as an hour. 
During each timestep, the independent system operator (ISO) of a wholesale electricity market collects aggregate supply and demand bids and solves an optimization problem to determine the least-cost dispatch that satisfies system constraints. In this paper, we let the ISO solve a simplified version of the economic dispatch (ED) problem (see detailed formulation in Appendix~\ref{app:ed}). The marginal cost of supplying one additional unit of electricity at each node, exactly the LMPs, is derived from the dual variables associated with the power balance and transmission constraints. We let $\lambda^n_t$ denote the LMP at bus $n \in \{1, \ldots, N\}$ and time $t$. For notation simplicity, let $\mD_t \coloneq (D^1_t, \cdots, D^N_t)$ denote the vector of all demand bids; then the LMP $\lambda^n_t$ is a function of $\mD_t$, which we write as $\lambda^n_t(\mD_t)$.



\section{Aggregators' Problem and the Mean-Field Framework}
\label{sec:Agg}
In this section, we first formulate the decision-making problem for each aggregator and then introduce the notion of an MFE. Acting on behalf of a large population of prosumers equipped with solar PVs and energy storage, each aggregator must determine control strategies for the prosumers under multiple sources of uncertainty, including solar energy output, real-time energy demand, and market prices. A key challenge lies in the presence of energy storage, which links decisions across time and leads to a high-dimensional stochastic dynamic optimization problem. To address this complexity in a scalable manner, we adopt an MFC formulation. In the infinite-population limit, the behavior of many similar prosumers converges to that of a representative agent, allowing the aggregator’s collective control problem to be expressed through this limiting agent. This approximation is standard in MFC and enables decentralized learning of control policies without tracking each individual prosumer. Importantly, the representative-agent approximation does not imply that all prosumers within an aggregator take identical actions. A policy is a mapping from a prosumer’s state to an action, and different prosumers can occupy different states even within the same aggregator. As a result, a single population-level policy can still generate heterogeneous behaviors across prosumers. Moreover, the policies we learn are stochastic, so even prosumers in similar states need not act identically. The precise definitions of states and policies will be introduced later in this section.
\subsection{Components of An Aggregator's Problem}

We begin by introducing the key components of the game. Throughout, we use $\mathcal{P}(\mathcal{X})$ for the set of Borel probability measures on a space $\mathcal{X}$.

\textbf{Time:} In power systems, many processes exhibit strong intra-day cycles that also repeat with similar patterns across days. For example, electricity demand and solar generation vary significantly across hours of the day, yet their daily profiles tend to follow recurring diurnal patterns. To capture both intra-day variation and inter-day similarity, we define two mappings over the global timestep \( t = 0, 1, 2, \ldots \). Let \( H \) denote the number of timesteps in a day, then
$
\Thour(t) = t \bmod H$, and  $\Tday(t) = \left\lfloor \frac{t}{H} \right\rfloor,
$
which return the hour within the day and the day index, respectively. These mappings enable the model to encode hourly dynamics while treating each day as structurally similar.

\textbf{Household agents (prosumers and consumers):}
At each bus $n$, there are $M^p_n$ prosumers, each equipped with rooftop PV and energy storage. The total aggregated storage capacity at bus $n$ is capped at $\overline{E}^n$. To capture heterogeneity, we partition the $M^p_n$ prosumers into $K$ types indexed by $k \in \{1,\ldots,K\}$, reflecting differences in storage capacity and associated PV size. Let $b^n_k \in [0,1]$ denote the fraction of type-$k$ prosumers at bus $n$, with $\sum_{k=1}^K b^n_k = 1$. Each type $k$ is assigned a relative (unnormalized) capacity parameter $\theta^n_k > 0$. For example, $\theta^n_2 = 2\theta^n_1$ implies that type-2 prosumers have twice the storage capacity of type-1 prosumers. The storage capacity of a type-$k$ prosumer $i$ at bus $n$ is then defined as
$
    \overline{E}^n_{ik}
    := \theta^n_k \overline{E}^n / M^p_n \sum_{\kappa=1}^K \theta^n_\kappa b^n_\kappa.
$
This scaling ensures that the total storage capacity at each bus remains finite in the mean-field limit, where the number of agents grows large and the impact of any individual agent becomes infinitesimal.

We express each prosumer's net demand in normalized form. For notational simplicity, we omit the type-$k$ subscript unless needed. At time $t$, let $\nd^n_{it}$ denote the ratio of prosumer $i$'s net demand to its storage capacity $\overline{E}^n_i$, so that $\nd^n_{it} \in [-1,1]$. Net demand is stochastic; we denote its (unknown) distribution by $\mathcal{Q}^p_n$. Neither prosumers nor aggregators are assumed to know $\mathcal{Q}^p_n$ a priori. Instead, decisions are  learned through repeated interaction with the environment.

At each bus $n$, there are also $M^c_n$ pure consumers. To enable a unified treatment, we assign each consumer a notional reference storage capacity $\overline{E}^n_j$ and define $\nd^n_{jt}$ as the ratio of demand to this reference capacity. This quantity is also stochastic, with distribution denoted by $\mathcal{Q}^c_n$.

\textbf{Aggregators:}
Although individual prosumers could in principle learn their own policies, this is impractical in practice, as most do not meet the 100~kW minimum threshold required by FERC Order No.~2222 for wholesale market participation. We therefore assume that, at each bus $n$, a single aggregator represents all prosumers at that location. While prosumers may differ in PV and storage capacities, they face the same LMPs and experience similar weather conditions at a given bus, which justifies modeling them with a common policy. We assume that PV generation is first used to meet local demand, with any surplus automatically charged into storage. The only decision variable is the storage action: how much energy to charge from or discharge to the market. Each aggregator learns a policy that maps system states (defined below) to charging or discharging actions. This policy is updated over time and broadcast to all prosumers at the bus, who use the policy to choose actions. Aggregators act as the learning agents, while prosumers are passive followers. As a starting point, aggregators in this work are modeled as non-strategic price takers rather than as entities exercising market power. Strategic behavior modeling and analysis of aggregators are deferred to future research.

\textbf{Actions:} As noted above, the only decisions in our setup involve charging or discharging energy storage. Let \( \mathcal{A} \subseteq [-1, 1] \) denote the action space for each aggregator, where each action \( a^n_t \in \mathcal{A} \) represents the proportion of storage capacity to charge (if \( a^n_t > 0 \)) or discharge (if \( a^n_t < 0 \)) at time \( t \), drawn according to the policy learned by the aggregator at bus \( n \) and executed by individual prosumers. To ensure feasibility, we implement \textit{action masking}, a common technique in RL that eliminates invalid actions based on the current state. 

\textbf{States:} In our model, each aggregator at bus $n$ is characterized by three state variables at time $t$: 
(i) storage level $e^n_{t} \in [0,1]$ (as a percentage of capacity), 
(ii) net load $\nd^n_{t}$ (unrelated to storage charging/discharging), and 
(iii) the current hour of the day, $\Thour(t)$. 
Let the state space be $\mathcal{S} \subseteq \RR^3$, where each element is a tuple of the form $s^n_{t} := (e^n_{t}, \nd^n_{t}, \Thour(t)) \in \mathcal{S}$.

Among the state variables, the net load $\nd^n_{t}$ is an exogenous random variable, mainly driven by solar irradiance and consumption behavior, and is unaffected by aggregators' actions. Time of day is deterministic. The only state variable affected by decisions is the storage level, which evolves according to the following rule after action $a^n_{t}$:
\begin{equation}
    e^n_{t+1} := \max \{\min \{e^n_{t} + a^n_{t}, 1\}, 0 \}.
    \label{eqn:storage-transition}
\end{equation}



\textbf{Mean Field:} With states and actions defined, the next step in standard RL (and Markov games) setups is to specify each agent's reward function. In our setting, however, rewards do not depend on the individual actions of other agents but only on the aggregate behavior of the population. Since we consider a large (infinite) number of agents, the influence of any single agent is negligible. Instead, each agent responds to the population-wide distribution of states and actions, known as the mean field. For simplicity and tractability, we begin by assuming that the state space \( \mathcal{S} \) and action space \( \mathcal{A} \) are discrete and finite. Let \( \mathcal{P}(\mathcal{S} \times \mathcal{A}) \) denote the space of probability distributions over the joint state–action space \( \mathcal{S} \times \mathcal{A} \); the formal definition of a mean field at bus $n$ and time $t$, 
\( \mathcal{L}^n_t \in \mathcal{P}(\mathcal{S} \times \mathcal{A}) \), is as follows: 
\begin{equation*}\label{eq:MF}
    \mathcal{L}^n_t(s, a) := \lim_{M^p_n \to \infty} \frac{1}{M^p_n} \sum_{i = 1}^{M^p_n} \mathbbm{1}_{(s^n_{it}, a^n_{it}) = (s, a)}, \quad \forall s \in \mathcal{S}, \ a \in \mathcal{A},
\end{equation*}
where \( \mathbbm{1}_{(s^n_{it}, a^n_{it}) = (s, a)} = 1 \) if prosumer \( i \) is in state \( s \) and takes action \( a \) at time \( t \), and 0 otherwise. Intuitively, the mean field is the limit of the histogram of joint state–action pairs as the number of prosumers goes to infinity. For notation simplicity, let $\vLL_t \coloneq (L^1_t, \cdots, L^N_t)$ denote the vector of all buses' mean-field; then our formulation allows us to equivalently treat the LMP as a function of $\vLL_t$, which we write as $\lambda^n_t(\vLL_t)$.

\textbf{Reward:} We now define the single-period reward function \( r_n: \mathcal{S} \times \mathcal{A} \times \RR^H \to \RR \) for each aggregator \( n \). Let \( s^n_t \) and \( a^n_t \) be the state and action at time \( t \), and let \( \vlambda^n := (\lambda^n_1, \dots, \lambda^n_H) \) denote the LMP profile at bus \( n \). The reward is given by
\begin{equation}
    r_n(s^n_t, a^n_t, \vlambda^n) = - \lambda^n_{\Thour(t)} \cdot \overline{E}_n \cdot \big( \Phi(e^n_t, a^n_t, \eta_n) + \nd^n_t \big),
    \label{eqn:rew}
\end{equation}
where \( \eta_n \in (0, 1] \) is the storage efficiency (assumed uniform across prosumers at bus \( n \)), and \( \Phi(\cdot) \) adjusts the action for efficiency losses:
\begin{equation}\label{eqn:adjusted-action}
   \Phi(e, a, \eta) = 
   \begin{cases}
   \max \{-e, a \} \cdot \eta, & \text{if } a < 0\ \text{(discharging)}, \\
   \min \{1 - e, a \} / \eta, & \text{if } a \geq 0\ \text{(charging)}. 
   \end{cases}
\end{equation}
The reward represents net profit, computed as negative price multiplied by total energy demand. 
To promote exploration and improve learning stability, we incorporate \textit{entropy regularization} into the reward function. Specifically, the single-period reward is modified as:
\begin{equation}\label{eq:r_Reg}
r^\reg_n(s, a, \vlambda^n) = r_n(s, a, \vlambda^n) - \Omega(\pi(\cdot \mid s, \vlambda^n)),
\end{equation}
where \( \Omega(\cdot) \) is a \( \rho \)-strongly convex regularization function. A common choice is the negative entropy:
\[
\Omega(\pi(\cdot \mid s, \vlambda^n)) = \alpha \sum_{a \in \mathcal{A}} \pi(a \mid s, \vlambda^n) \log \pi(a \mid s, \vlambda^n),
\]
with \( \alpha > 0 \) controlling the strength of regularization. 

\textbf{Policy:}  
At each decision time \( t \), aggregator \( n \) learns a policy  $
\pi^n_t: \mathcal{S} \times \RR^H \to \mathcal{P}(\mathcal{A}),$ 
which maps each state and LMP profile to a distribution over actions. Given the LMP profile \( \vlambda^n \in \RR^H \) and initial state \( s^n_0 \), the \textit{regularized value function} under policy \( \pi^n_t \) is defined as the expected sum of discounted rewards:
\begin{equation}\label{eq:V_Reg}
V^\reg_n(s^n_0, \pi^n_t, \vlambda^n) = \bbE \left[ \sum_{\tau=0}^{\infty} \gamma_n^\tau r^\reg_n(s^n_{\tau}, a^n_{\tau}, \vlambda^n) \right],
\end{equation}
where \( \gamma_n \in (0, 1] \) is the discount factor, and the expectation is over the trajectory induced by the policy and external uncertainties. Each aggregator seeks to learn an optimal policy \( \pi^{n*}_t \) that maximizes \( V^\reg_n \) at each time \( t \).

\subsection{Mean Field Equilibrium} \label{subsec:mfe}
An MFE is a fixed point in a dynamic system of infinitely many interacting agents, characterized by two key properties: (i) \textbf{optimality}, where each agent's strategy is optimal given the mean field, and (ii) \textbf{consistency}, where the mean field coincides with the distribution of states and actions induced by all agents following that strategy. Let $\Pi := \left\{ \pi \mid \pi: \mathcal{S} \times \RR^H \to \mathcal{P}(\mathcal{A}) \right\}$ denote the space of state–dependent stochastic policies. Let $\vpi = (\pi^{1}, \cdots, \pi^{N})$ denote the policy profile of all aggregators. The formal definition of an MFE is given as follows:
\begin{definition}[MFE]
A profile $\{(\pi^{n*}, \vlambda^{n*})\}_{n=1}^{N}$ is a mean field equilibrium if, for each aggregator $n$ and any state $s^n \in \mathcal{S}$ at time $t$, the following conditions hold:
\begin{enumerate}
    \item \textbf{Optimality:} Given $\vlambda^{n*}$, each aggregator finds the optimal policy $\pi^{n*}$, such that for any policy \( \pi^n \in \Pi \),
    \[
    V^\reg_n(s^n, \pi^{n*}, \vlambda^n) \geq V^\reg_n(s^n, \pi^n, \vlambda^n).
    \]
    
    \item \textbf{Consistency:} If all aggregators follow $\vpi^*$, then the mean field distribution evolves according to: $\LL^n_{t+1} = \Gamma^n(\LL^n_t, \pi^{n*}),$ where $\Gamma^n$ is the consistency operator defined in the following form:
    \begin{align}
        & \Gamma^n(\LL^n, \pi^n)(s', a') := \zeta \frac{1}{\abs{\mathcal{S}}\abs{\mathcal{A}}} \nonumber \\
        & \hspace{10pt} + (1 - \zeta)\sum_{s, a}\LL^n(s, a) \cdot P^n(s' \mid s, a') \cdot \pi^n(a' \mid s, \vlambda),
        \label{eq:update-mf}
    \end{align}
    for all $s' \in \mathcal{S}, a' \in \mathcal{A}$, and $\zeta \in (0, 1)$ is the probability of a uniform noise into the MF update.  \( P^n(s' \mid s, a') \) denotes the state transition probability for the aggregator at bus \( n \). 
    The corresponding LMP obtained through the ED problem \eqref{eqn:ed-obj} -- \eqref{eqn:gen-limit}, 
    $\lambda^n_{t+1}(\vLL_{t+1})$, must satisfy $\lambda^n_{t+1}(\vLL_{t+1}) = \left[\vlambda^{n*}\right]_{\Thour(t+1)}.$
\end{enumerate}
\end{definition}
\noindent The parameter $\zeta$ in \eqref{eq:update-mf} is a regeneration probability that gives each prosumer a small chance of resetting to a uniformly sampled state. Economically, this models prosumer turnover at a bus, new prosumers entering and existing ones exiting, and keeps the environment dynamic even at steady state. This regeneration mechanism prevents the mean field from becoming static and supports continual learning.

We establish in Appendix~\ref{app:proof} the conditions under which such an MFE exists and is unique.

Given the definition above, a natural fixed-point iteration for approximating an MFE proceeds as follows:  
(i) Fix $\LL^n_t$ and $\vlambda^n_t$. For a given state $s^n$, aggregator $n$ computes its optimal policy $\pi_t^{n*}$.  
(ii) Update the mean field using $\LL^n_{t+1} = \Gamma^n(\LL^n_t, \pi_t^{n*})$ and compute the updated LMP $\lambda^n_{t+1}(\vLL_{t+1})$ by solving the ED problem \eqref{eqn:ed-obj}--\eqref{eqn:gen-limit}. If $t+1$ marks the start of a new day, update $\vlambda^n$ using the most recent $H$ LMPs.  
(iii) Repeat steps (i)--(ii) until $\vlambda^n_{t+1}$ stabilizes.

The key idea behind this framework is that information about the mean field is embedded in the LMPs. Aggregators update their policies based on observed LMPs, and the mean field evolves in response to these updated policies. We also note that an MFE is not generally a Nash equilibrium, as the latter is typically defined for games with a finite number of agents. Whether an MFE arises as the limit of an \( M \)-agent Nash equilibrium as \( M\to \infty \) is a nontrivial question. \cite{Basar_MFE_Nash} provides sufficient conditions under which the infinite-agent MFE policy yields an \( \epsilon \)-Nash equilibrium for the corresponding finite-\( M \) game. In this work, we focus on the properties of the MFE in the infinite-agent setting and leave its connection to finite-agent Nash equilibria to future research.
\section{The Two-Phase RL Algorithm}
\label{sec:alg}

We propose a two-phase, distributed mean-field RL algorithm executed at each time period $t$. 
\textbf{Phase 1 (Training):} each aggregator independently trains its policy using an RL algorithm. \textbf{Phase 2 (Execution):} prosumers act according to their aggregator’s trained policy. Since policies are stochastic, identical prosumers may take different actions. These actions are aggregated into bids and submitted to the market clearing process, after which LMPs for time period $t$ are determined by solving the ED problem. 

Aggregators do not have direct knowledge of external uncertainty distributions, such as the net load. However, an important feature of electricity markets is that LMP fluctuations implicitly reflect both market dynamics and exogenous uncertainties. Following the approach in~\cite{feng2025decentralized}, we allow each aggregator to maintain a belief over the LMPs for each hour of the day, represented by a vector of length $H$. These beliefs are formed from historical observations and guide the RL-based policy training in Phase~1. After market clearing, aggregators update their beliefs using the realized LMPs.

All prosumers at the same bus share the belief vector maintained by their aggregator, since they face identical LMPs. Let $\vbelief^n_t \in \mathbb{R}^H$ denote the belief vector of aggregator $n$ at time $t$. Once the ISO solves the ED problem and returns LMP $\lambda^n_t$ at bus $n$, the aggregator updates its belief as
\begin{equation}
    \vbelief^n_{t+1}
    := \vbelief^n_t
    - \frac{\delta_n}{\sqrt{\Tday(t) + 1}}
    \left((\vbelief^n_t)^\top \mathbf{1}_{\Thour(t)} - \lambda^n_t\right)
    \mathbf{1}_{\Thour(t)},
    \label{eqn:lmp-belief}
\end{equation}
where $\delta_n \in [0.5,1]$ is a learning-rate parameter and $\mathbf{1}_{\Thour(t)} \in \mathbb{R}^H$ is the unit vector with a 1 at entry $\Thour(t)$ and zeros elsewhere. Thus, at each time $t$, only the belief corresponding to hour $\Thour(t)$ is updated.

\textbf{Training Phase:} At the beginning of each time period $t$, each aggregator fixes its LMP belief and trains a policy using an RL algorithm, referred to generically as \textit{Alg} (such as PPO, TRPO, or SAC), for $\Ttrain$ steps.

\textbf{Execution Phase:} After each aggregator has learned a policy, this policy is distributed to its prosumers. At bus $n$, prosumer $i$ takes an action from the policy $a^n_{it} \sim \pi^{n*}_t(\cdot | s^n_{it}, \belief^n_t)$. Also, each prosumer $i$ and consumer $j$'s original net demand are realized following the distributions $\mathcal{Q}^p_n$ and $\mathcal{Q}^c_n$, respectively. The net demand in the quantity of energy now has the following forms for prosumer $i$ at bus $n$ at time $t$:
   $ d^n_{it} = (\Phi(e^n_{t}, a^n_{t}, \eta_n) + \nd^n_{t}) \overline{E}^n_i,
$
and similarly, for each consumer $j$ at bus $n$ at time $t$:
$
    d^n_{jt} = \nd^n_{jt} \overline{E}^n_j.
$

The transition of the state of charge for each prosumer $i$ at bus $n$ from time $t$ to $t+1$ is defined as follows: 
\begin{equation}
    e^n_{i,t+1} = \begin{cases}
        \text{Uniform}(0, 1), \text{with probability $\zeta$}, \\
        \max \left\{\min \left\{e^n_{it} + a^n_{it}, 1\right\}, 0 \right\}, \text{with probability $1-\zeta$},
    \end{cases}
    \label{eqn:random-regen}
\end{equation}
where $\zeta$ is the regeneration probability introduced in \eqref{eq:update-mf}. 
The aggregator at $n$ updates its storage level as:
\begin{equation}
    e^n_{t+1} = \frac{1}{\overline{E}_n}\sum_{i=1}^{M^p_n} e^n_{i,t+1} \overline{E}^n_i,
    \label{eqn:agg-soc-transition}
\end{equation}
which is a weighted average of all its prosumers storage level. This storage level is then used as the initial storage level state to begin the RL training for time $t$ for $\Ttrain$ steps. The pseudo-code is presented in Algorithm \ref{alg:two-phase} in Appendix~\ref{app:alg}. 

It is important to note that the RL training in Phase~1 is performed offline and does not operate on real-time system timescales. Although historical LMPs are used to initialize each aggregator’s belief vector, the training itself does not rely on historical prices. Instead, during offline training the aggregator interacts with a simulated environment in which LMPs are generated endogenously by solving the ED model in response to the aggregator’s actions and the evolving mean field. This preserves the closed-loop feedback between aggregator behavior and market prices while avoiding any real-time computational burden. Aggregators periodically update their policies by retraining them offline using newly observed LMPs, updated belief vectors, and accumulated historical trajectories, and these computations can be carried out on dedicated cloud or local servers without interfering with market operations. Once an updated policy is obtained, it is deployed in real time by simply broadcasting it to prosumers. During real-time operation, only Phase~2 is executed, and this phase requires negligible computation. Thus, the proposed framework does not impose any real-time computational burden.

\textit{Remark on distribution network constraints.} 
The baseline formulation models aggregators at transmission-level buses and therefore does not explicitly include distribution network constraints. However, such constraints can be incorporated into our framework without modifying its overall structure. In particular, voltage or line-flow limits at the distribution level can be enforced through negative reward terms during RL training. This approach follows the same principle used in peer-to-peer energy trading in distribution networks \cite{P2P_FengLiu}. Such extensions are not pursued in this work, as our focus is on presenting the framework for wholesale-level market interactions.

\section{Numerical Experiment}
\label{sec:numerical}
We evaluate the proposed framework on the 37-bus synthetic Oahu network from \cite{birchfield}. Each Hawaiian Electric generating plant is mapped to its nearest bus using data from \cite{powerfacts}, resulting in 26 generators: 4 oil, 2 biomass, 17 utility-scale solar (distinct from household PV), and 3 wind units. Oil and biomass units use quadratic costs \(C(p)=ap^{2}+bp\), where \(p\) is MW and \(C(p)\) is \$/h, with parameter ranges from \cite{isone, tidball2010cost}. For oil, \(a \in [0.0059,0.0342]\,\$/\mathrm{MW}^{2}\mathrm{h}\) and \(b=19.98\,\$/\mathrm{MWh}\); for biomass, \(a \in [0.001,0.002]\,\$/\mathrm{MW}^{2}\mathrm{h}\) and \(b \in [28.45,52.65]\,\$/\mathrm{MWh}\), sampled uniformly. Solar and wind have zero marginal cost. Their capacity factors follow historical hourly profiles, solar from \cite{profilesolar}, wind from \cite{argueso}, and are multiplied by triangular noise: \(\Delta(0.8,1.2,1)\) for solar and \(\Delta(0.5,1.5,1)\) for wind.

Each bus hosts 650 prosumers and 2{,}000 consumers. Prosumers consist of 500 small, 100 medium, and 50 large types with storage capacities of 10, 20, and 30~kWh. Demand profiles follow \cite{coffman2016estimating}. Larger storage capacities imply proportionally scaled net-load levels in the dataset. To introduce uncertainty, each agent’s mean daily demand is scaled by \(\Delta(0.8,1.2,1)\).

We implement the PPO RL algorithm with \(T_{\text{train}}=1{,}200\), \(H=12\) (2-hour intervals), \(\delta_n=0.9\), and \(\zeta=0.01\). Each simulation runs 50 days and is repeated five times using different seeds. Experiments are run on a Windows 11 machine with an Intel i7 CPU and an NVIDIA RTX 4070 GPU.

Figure~\ref{fig:hub-prices} shows hub prices over the first and last 3 days of training. We compare 3 scenarios: the proposed MFE-based framework, a decentralized heuristic algorithm (DHA; see Appendix~\ref{app:dha}), and a baseline without storage (grid-tied PV only). Initially, MFE prices behave similarly to the no-storage case, while DHA exhibits the lowest volatility. As learning progresses, prices under MFE become less volatile and converge to a stable daily pattern, indicating convergence to a steady state.

\begin{figure}[!ht]
    \begin{center}
        \begin{tabular}{@{}c@{}}
          \centering
          \includegraphics[width=0.9\columnwidth]{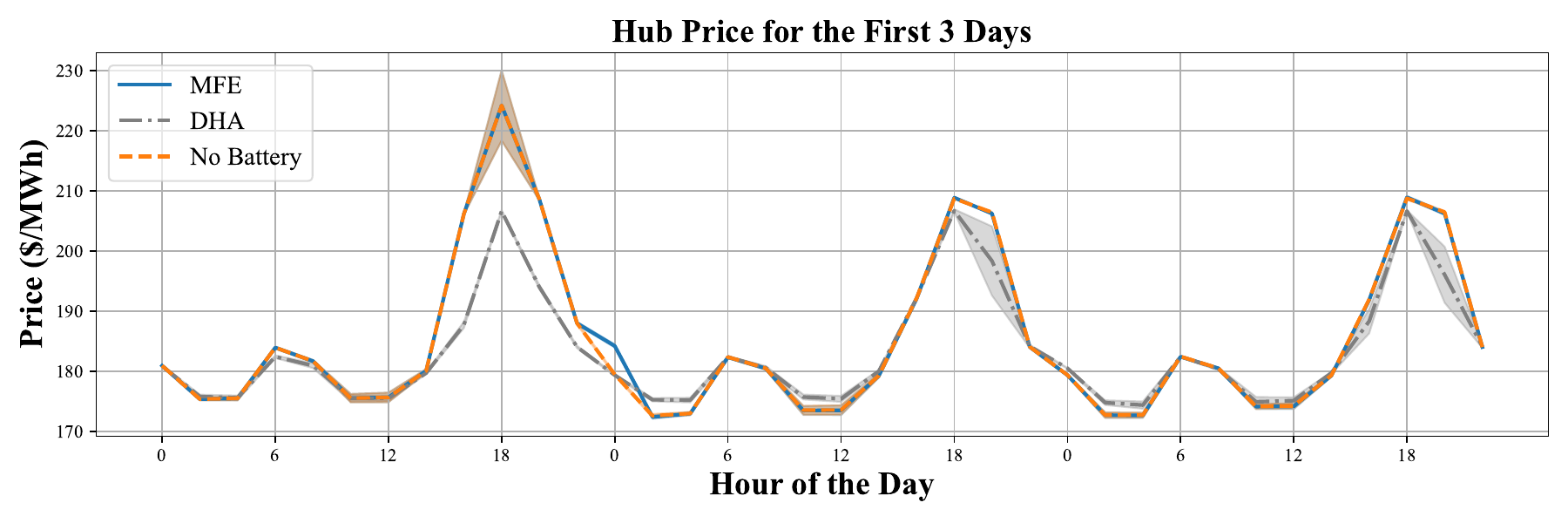}  
          \label{fig:hub-prices-first}
        \end{tabular}
        \begin{tabular}{@{}c@{}}
          \centering
          \includegraphics[width=0.9\columnwidth]{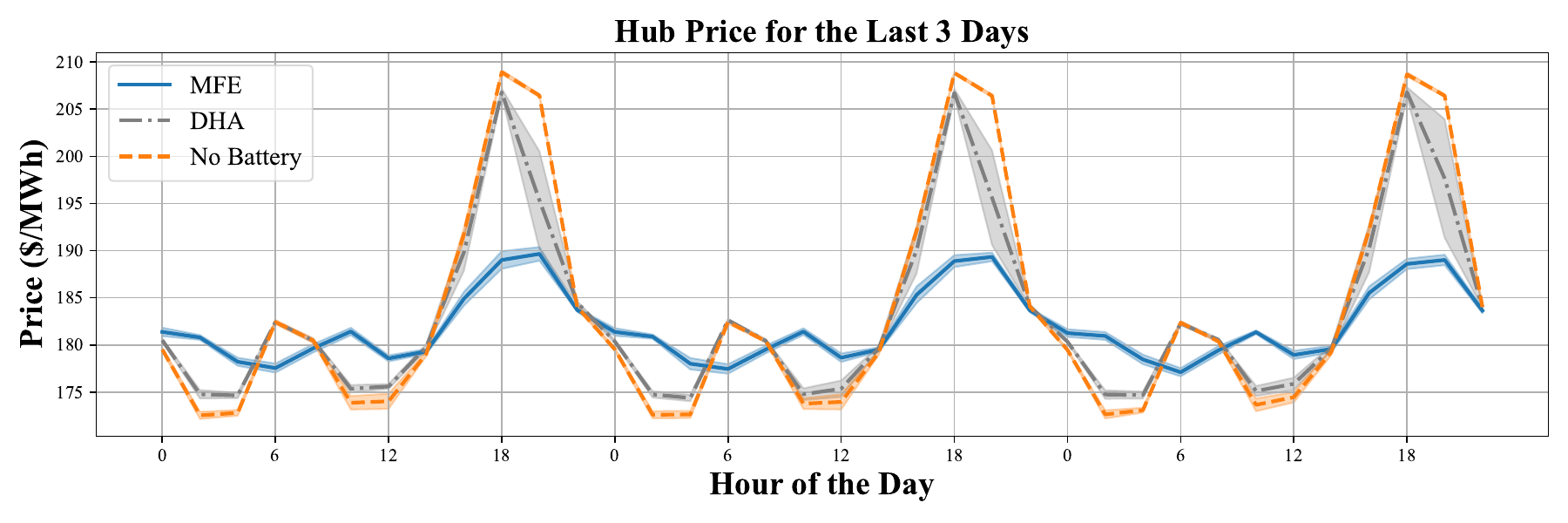}  
          \label{fig:hub-prices-last}
        \end{tabular}
        \caption{LMP hub price Comparisons (shaded areas show one-standard-deviation error bounds across simulations).}
        \label{fig:hub-prices}
    \end{center}
\end{figure}

To quantify price volatility, we adopt the incremental mean volatility (IMV) metric from \cite{roozbehani}:
$
  \mathrm{IMV} = \lim_{T \to \infty} \frac{1}{T} \sum_{t=1}^T \abs{\lambda_{t+1}-\lambda_t}.
$ Figure~\ref{fig:imv} reports IMV values over the final 3 days. The MFE scenario yields significantly lower IMV, indicating reduced price volatility.

\begin{figure}[!ht]
    \centering
    \includegraphics[width=0.9\columnwidth]{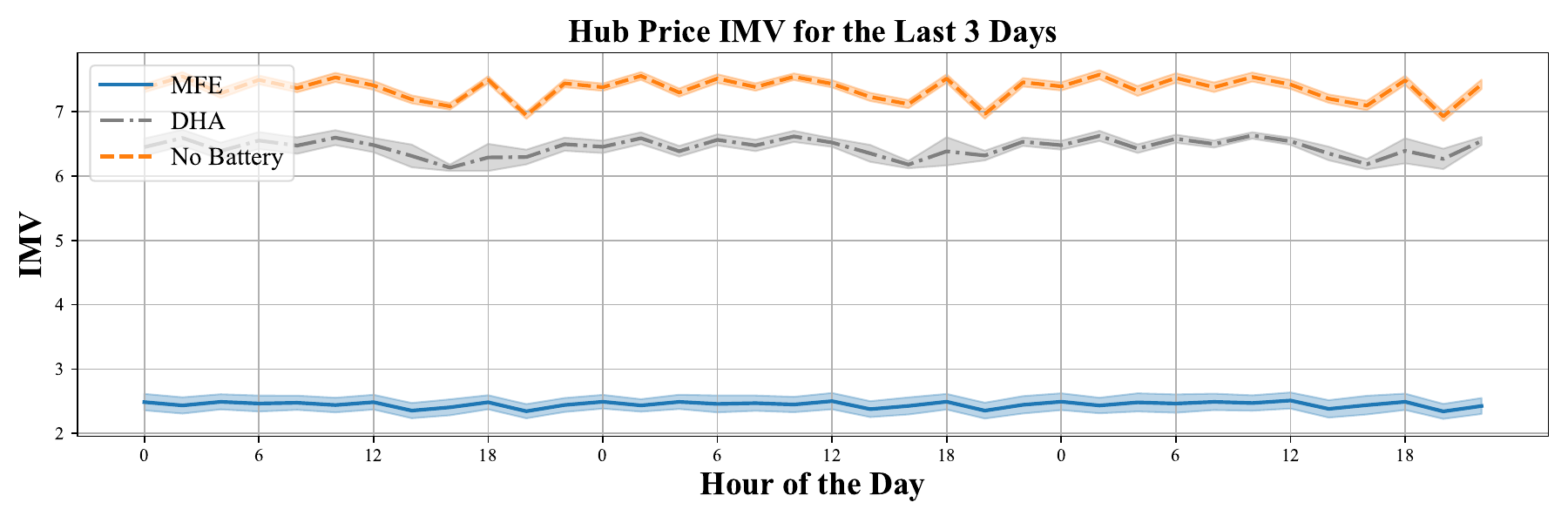}  
    \caption{Comparison of IMV over the last 3 days across scenarios. Shaded areas show one-standard-deviation error bounds across simulations.}
    \label{fig:imv}
\end{figure}

We also compute the \textit{ex-post} daily cost for prosumers and consumers, defined as the sum over time of realized LMPs multiplied by cleared energy quantities. Figure~\ref{fig:cost} shows that total daily costs are lowest under MFE for both groups.

\begin{figure}[!ht]
    \begin{center}
        \begin{tabular}{@{}c@{}}
          \centering
          \includegraphics[width=0.9\columnwidth]{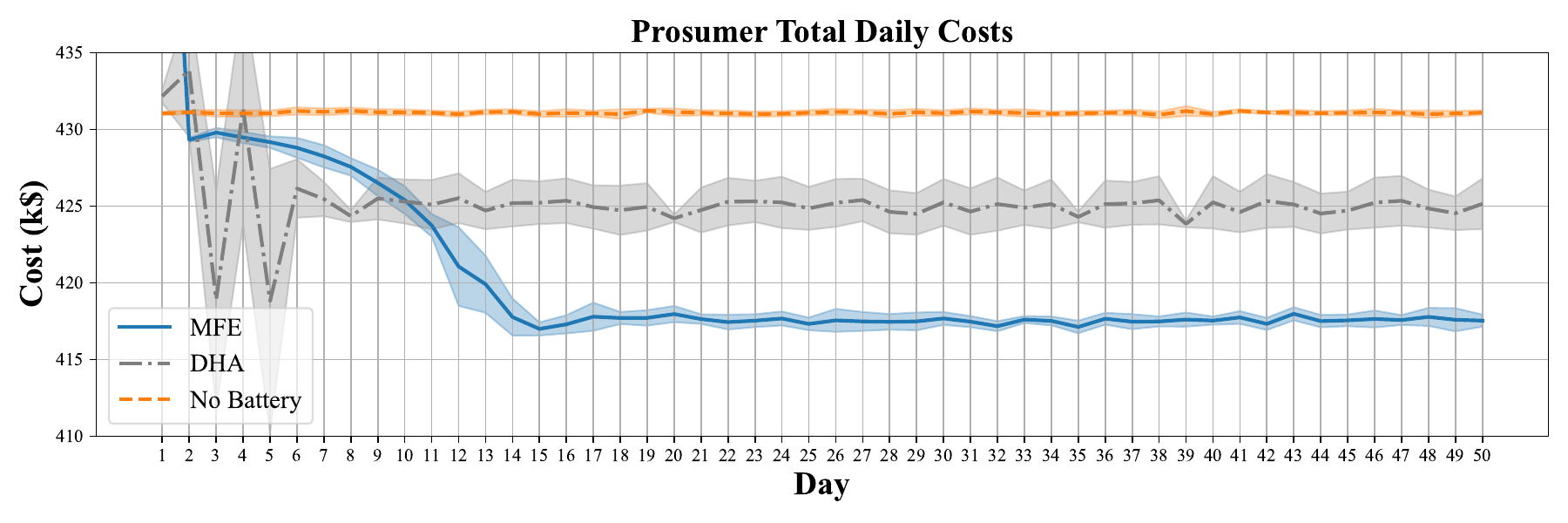}  
          \label{fig:cost-prosumer}
        \end{tabular}
        \begin{tabular}{@{}c@{}}
          \centering
          \includegraphics[width=0.9\columnwidth]{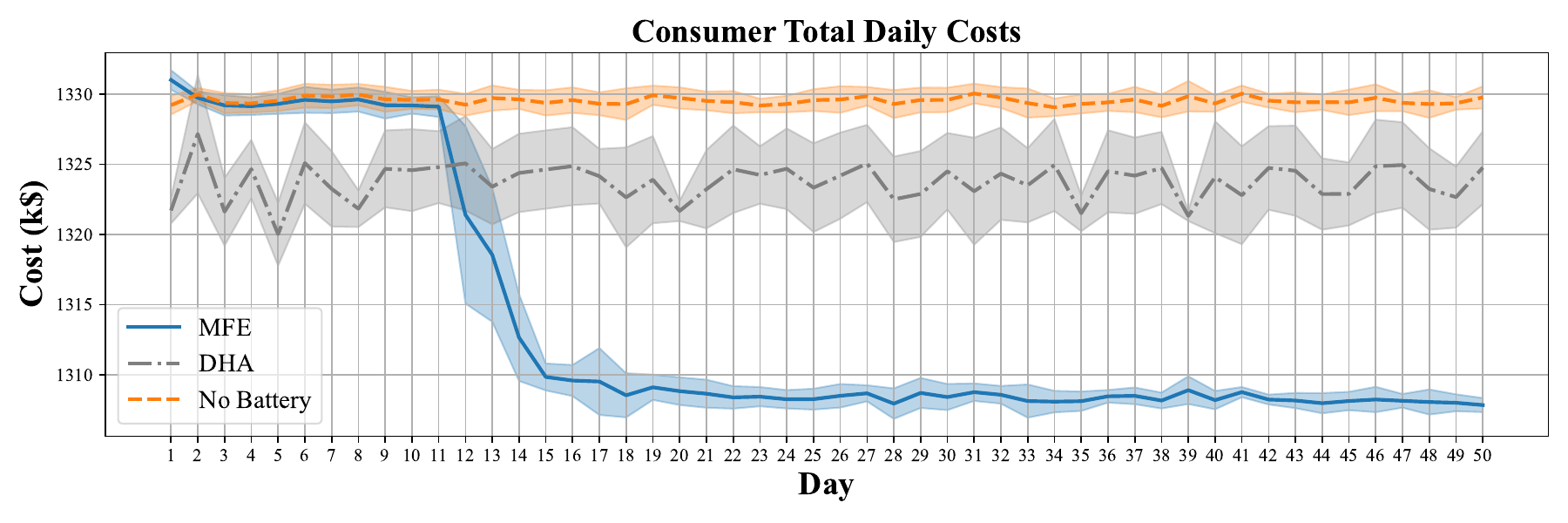}  
          \label{fig:cost-consumer}
        \end{tabular}
        \caption{Comparison of daily costs for prosumers and consumers. Shaded areas show one-standard-deviation error bounds across simulations.}
        \label{fig:cost}
    \end{center}
\end{figure}

Finally, we examine the impact on the “duck curve,” that is, the net demand curves obtained by subtracting solar generation from the gross system load. 
Figure~\ref{fig:final-shape} compares the net demand under the no-storage, DHA, and MFE cases, averaged over the final ten days. 
The MFE policy produces the greatest degree of load shifting, charging during periods of abundant midday sunshine and reducing the evening peak, thereby smoothing the aggregate load profile.

\begin{figure}[!ht]
    \centering
    \includegraphics[width=0.9\columnwidth]{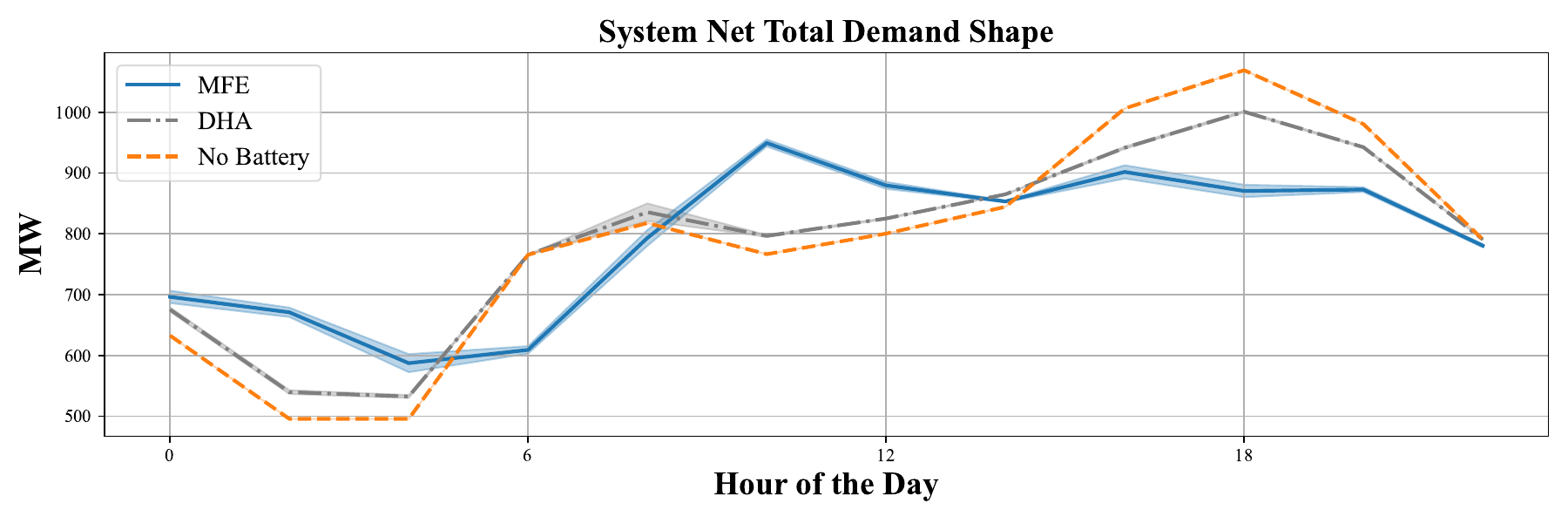}  
    \caption{Comparison of net load curves.}
    \label{fig:final-shape}
\end{figure}
\section{Conclusion} \label{sec:conclusion}
This paper develops an algorithmic framework for integrating DERs into wholesale electricity markets through decentralized, RL-enabled aggregators. Aggregators learn storage charging and discharging strategies for prosumers under uncertainty in renewable output, demand, and market prices. A central idea is to treat LMPs as a mean-field signal that reflects aggregate supply-demand interactions and network constraints. Within this hybrid MFC–MFE setting, we propose a two-phase RL algorithm that enables fully decentralized policy learning.

We also establish sufficient conditions for the existence and uniqueness of an MFE and show convergence of the resulting three-step learning process in the infinite-agent limit. Numerical experiments demonstrate that coordinated storage control within this framework substantially reduces LMP volatility and lowers costs for both prosumers and consumers, highlighting the value of market-integrated DER coordination.

Future work will proceed in three directions. First, although Algorithm~1 provides a natural fixed-point–style approximation of an MFE, each policy update requires an evolving LMP belief, making the method heuristic in practice. An important direction is to analyze the belief-update dynamics and determine conditions under which LMP beliefs, and consequently Algorithm~1, converge. Second, we aim to characterize the relationship between the infinite-population MFE and the Markov–Nash equilibrium of a finite but large number of aggregators, identifying regimes where the latter converges to the former. Third, we plan to incorporate strategic aggregators who internalize their price impact, enabling a systematic study of market power and strategic interactions within the mean-field framework.

\section*{Acknowledgements}
The authors used ChatGPT solely for language polishing and editorial refinement. All research ideas, modeling, analysis, and results were developed independently by the authors.

\appendices
\section{Economic Dispatch (ED)} \label{app:ed}
To provide the specific formula of LMPs, we present a basic economic dispatch (ED) problem solved by an ISO. Although actual dispatch problems are considerably more complex, this  formulation retains the essential constraints: power balance, generation capacity, and transmission limits, and is sufficient to demonstrate how prices are formed and how they relate to system demand. Importantly, our framework is not restricted to this simplification. A key strength of the framework is that it requires no modification to ISO operations: system operators can continue using their full-scale market-clearing processes without adjustment. All learning and coordination occur at the aggregator level, ensuring full compatibility with existing market structures. The simplified ED problem is given as follows:
\begin{align}
    & \min_{p_{1t}, \cdots, p_{Gt}} \; \sum_{g=1}^G C_g(p_{gt}) \label{eqn:ed-obj} \\
    & \text{s.t.} \; \sum_{g=1}^G p_{gt} = \sum_{n=1}^{N} D^n_t, \quad (\lambda^{\text{HUB}}_t) \label{eqn:power-balance} \\
    & \quad -\overline{F}_l \leq \sum_{n=1}^{N} \text{PTDF}_{ln} \left( \sum_{g \in G_n} p_{gt} - D^n_t \right) \leq \overline{F}_l,\ (\underline{\mu}_{lt}, \overline{\mu}_{lt}) \nonumber \\
    & \hspace{125pt}\quad \forall \; l \in \{1, \ldots, L\}, \label{eqn:flow-limit} \\
    & \quad 0 \leq p_{gt} \leq \overline{p}_g, \quad (\underline{\nu}_{gt}, \overline{\nu}_{gt}) \nonumber \\
    & \hspace{125pt}\quad \forall \; g \in \{1, \ldots, G\}. \label{eqn:gen-limit}
\end{align}
Here, the variables $p_{gt}$ represent the average power output of generator $g$ in time interval $t$. The parameter $\text{PTDF}_{ln}$ denotes the power transfer distribution factor for line $l$ and bus $n$, which reflects how power flows across the network; $\overline{F}_l$ represents the maximum allowable power flow on line $l$, and $\overline{p}_g$ represents the generation capacity of generator $g$. In addition, the dual variables are listed on the right-hand-side of each constraint. The dual of the power balance constraint \eqref{eqn:power-balance}, $\lambda^{\text{HUB}}_t$, represents the so-called hub price, and $\underline{\mu}_{lt}, \overline{\mu}_{lt}, \underline{\nu}_{gt}, \overline{\nu}_{gt}$ are the dual variables corresponding to line flow limits of each line $l$, and generator output limits of each generator $g$, respectively. The LMP at bus $n \in \{1, \ldots, N\}$ and time $t$, denoted $\lambda^n_t$, is derived as follows:
\begin{align}
    \lambda^n_t & \coloneq \frac{\partial \mathfrak{L}_t}{\partial D^n_t} = \lambda^{\text{HUB}}_t - \sum_{l=1}^L \text{PTDF}_{ln} (\underline{\mu}_{lt} - \overline{\mu}_{lt}), 
    \label{eqn:lmp}
\end{align}
where $\mathfrak{L}_t$ represents the Lagrangian function of the ED problem. 

\section{Existence and Uniqueness of an MFE}
\label{app:proof}
In establishing the existence and uniqueness of an MFE, a key technical condition needed is the Lipschitz continuity of the mapping \( \lambda^n_t(\mD_t) \).  The following result, established in \cite{feng2025decentralized}, provides sufficient conditions under which this property holds. For completeness, we restate it here, along with the required constraint qualification assumption.

\begin{assumption} (LICQ) \label{assump:LICQ}
    Let \( X(\mD_t) \) denote the feasible region of the economic dispatch problem defined in \eqref{eqn:ed-obj}–\eqref{eqn:gen-limit}. We assume that, for all \( t \) and all \( \mD_t \) such that \( X(\mD_t) \) is non-empty, the linear independence constraint qualification (LICQ) holds at every feasible point in \( X(\mD_t) \).
\end{assumption}

\begin{proposition} (Lipschitz Continuity of LMPs)\cite{feng2025decentralized} 
    Given Assumption \ref{assump:LICQ} and that each generator’s cost function \( C_g(\cdot) \) is strongly convex and quadratic. Then, for all buses \( n = 1, \ldots, N \), the LMP \( \lambda^n_t(\mD_t) \) is single-valued and Lipschitz continuous with respect to \( \mD_t \); that is, there exists a constant \( L_{\lambda} > 0 \) such that for any $\mD_t$ and $\widetilde{\mD}_t \geq 0$, $
        \abs{\lambda^n_t(\mD_t) - \lambda^n_{t}(\widetilde{\mD}_t)} \leq L_{\lambda} \normof{\mD_t - \widetilde{\mD}_t}.
$
    \label{prop:feng2025}
\end{proposition}

We now prove that the proposed 3-step framework in Section~\ref{subsec:mfe} defines a contraction mapping over the space of policy and price profiles, guaranteeing convergence to a unique fixed point. This fixed point corresponds exactly to an MFE, thereby establishing its existence and uniqueness. Unless otherwise noted, all norms here are $\ell$-norms.
 
\begin{theorem}[Lipschitz Optimal Solution]
    Under the same assumptions of Proposition \ref{prop:feng2025}, in the regularized MFG, for any belief $\vlambda^n, \vlambda^{n'}$, there exists $L \geq 0$ such that:
    \begin{equation}
        \sup_{s \in \mathcal{S}} \normof{\pi^{n*}(s, \vlambda^n) - \pi^{n*}(s, \vlambda^{n'})} \leq L \normof{\vlambda^n - \vlambda^{n'}}.
        \label{eq:lip-gamma1}
    \end{equation}
    \label{thm:op}
\end{theorem}
\begin{proof}[Proof Sketch]
We first note that under regularization, $V^\reg$ is a strongly concave function and therefore admits unique optimal solution. We then proceed in 3 steps. 
(i) For a fixed policy $\pi$, the difference between the optimal $Q$-functions $Q^*_{\vlambda}$ and $Q^*_{\vlambda_2}$ are bounded by $L_r \|\vlambda - \vlambda_2\|$, due to the Lipschitz continuity of the reward function. This implies that $Q^*_{\vlambda}$ is Lipschitz continuous with respect to $\vlambda$.
(ii) Define $q^s_{\vlambda} := r(s,\cdot,\vlambda) + \gamma \sum_{s'} Q^*_{\vlambda}(s') P(s'|s,\cdot)$, the action–value vector before regularization. This inherits the Lipschitz property with constant $\big(L_r+\tfrac{\gamma L_r}{1-\gamma}\big)$. 
(iii) By Fenchel duality, the optimal policy is given by $\pi^{n*}(s,\vlambda) = \nabla \Omega^*(q^s_{\vlambda})$, where $\Omega^*$ is the conjugate of the $\rho$-strongly convex regularizer $\Omega$. Since $\Omega^*$ is $1/\rho$-smooth, $\pi^{n*}$ is Lipschitz with constant $L=\tfrac{1}{\rho}\big(L_r+\tfrac{\gamma L_r}{1-\gamma}\big)$.  
\end{proof}

\begin{theorem}[Lipschitz Continuity of $\Gamma^n$]
    Under the same assumptions of Proposition \ref{prop:feng2025}, for any MF $\LL^n, \LL^{n'}$ and policies $\pi^n, \pi^{n'}$, given the belief $\vlambda^n$, there exists $L_2, L_3 \geq 0$ such that
    \begin{align}
        & \normof{\Gamma^n(\LL^n, \pi^n) - \Gamma^n(\LL^{n'}, \pi^n)} \leq L_2 \normof{\LL^n - \LL^{n'}}, \\
        & \normof{\Gamma^n(\LL^n, \pi^n) - \Gamma^n(\LL^n, \pi^{n'})} \nonumber \\
        & \hspace{40pt} \leq L_3 \sup_{s \in \mathcal{S}} \normof{\pi^n(s, \vlambda^n) - \pi^{n'}(s, \vlambda^n)}.
    \end{align}
    \label{thm:gamma2}
\end{theorem}
\begin{proof}[Proof Sketch]
We expand the no-regeneration transition operator as
\[
\tilde{\Gamma}^n(\LL^n,\pi^n)(s',a') = \sum_{s,a} \LL^n(s,a)\, P^n(s'|s,a)\, \pi^n(a'|s).
\]
Fixing $\pi^n$, the difference $\tilde{\Gamma}^n(\LL^n,\pi^n) - \tilde{\Gamma}^n(\LL^{n'},\pi^n)$ can be bounded linearly in $\|\LL^n - \LL^{n'}\|$ since the sum is a linear operator in $\LL^n$ with probability weights. Similarly, fixing $\LL^n$, the difference in $\pi^n$ can be bounded by $\sup_s \|\pi^n(\cdot|s) - \pi^{n'}(\cdot|s)\|$. Including regeneration with probability $\zeta$ scales both terms by $(1 - \zeta)$, yielding Lipschitz constants $L_2 = L_3 = 1 - \zeta$.
\end{proof}

We now present the main theorem regarding the MFE existence and uniqueness.
\begin{theorem}[Existence and Uniqueness of MFE]
    Under the same LICQ and strongly convex quadratic cost function assumptions as in Proposition \ref{prop:feng2025}, given that $L L_{\text{MF}} L_3 + L_2 < 1$, there exists a unique MFE following the 3-step procedure.
    \label{thm:mfe}
\end{theorem}
\begin{proof}[Proof Sketch]
We first show that LMPs are Lipschitz continuous with respect to the mean-field (MF) profile $\vLL_t := (\LL^1_t, \ldots, \LL^N_t)$. Fixing the consumer demand, the total demand at each bus $n$ is a weighted sum over $\LL^n_t(s,a)$ scaled by the storage rating $\overline{E}_n$, implying
\[
\|\mD_t - \mD_{t'}\|_1 \le \max_n \overline{E}_n \|\vLL_t - \vLL_{t'}\|_1.
\]
By Proposition~\ref{prop:feng2025}, LMPs are Lipschitz in total demand with constant $L_\lambda$. Combining the two gives
\[
|\lambda^n_t - \lambda^n_{t'}| \le L_{\text{MF}} \|\vLL_t - \vLL_{t'}\|_1, \quad \text{with } L_{\text{MF}} := L_\lambda \max_n \overline{E}_n.
\]

Now consider the update from time $t$ to $t+H$ (same hour on consecutive days). We then have:
\begin{align*}
& \|\Gamma^n(\LL_t^n, \pi^{n*}(s^n, \vlambda_t^n)) - \Gamma^n(\LL_{t+H}^n, \pi^{n*}(s^n, \vlambda_{t+H}^n))\|_1 \\
&\le L_3 \|\pi^{n*}(s^n, \vlambda_t^n) - \pi^{n*}(s^n, \vlambda_{t+H}^n)\|_1 + L_2 \|\LL_t^n - \LL_{t+H}^n\|_1 \\
&\le L_1 L_3 \|\vlambda_t^n - \vlambda_{t+H}^n\|_1 + L_2 \|\LL_t^n - \LL_{t+H}^n\|_1 \\
&\le L_1 L_{\text{MF}} L_3 \|\LL_t^n - \LL_{t+H}^n\|_1 + L_2 \|\LL_t^n - \LL_{t+H}^n\|_1.
\end{align*}
Thus, the full update operator is a contraction with constant $(L_1 L_{\text{MF}} L_3 + L_2)$ under the $\ell_1$ norm. If this constant is strictly less than 1, Banach’s fixed-point theorem guarantees existence and uniqueness of the MFE.
\end{proof}

\section{The Two-Phase RL Algorithm} \label{app:alg}
We present the pseudo-code of the two-phase RL algorithm in Algorithm~\ref{alg:two-phase} based on Section~\ref{sec:alg}.
\begin{algorithm}
    \caption{A two-phase distributed mean-field RL algorithm with LMP beliefs and entropy regularization}
    \label{alg:two-phase}
    \KwIn{Initial battery states $e^n_0 \in [0, 1]$, initial LMP beliefs $\vbelief^n_0 \in \mathbb{R}^H$ with learning rates $\delta_n \in [0.5, 1]$, demand shapes $\mathcal{Q}^p_n, \mathcal{Q}^c_n$; training step $\Ttrain$; random regeneration probability $\zeta$; an RL algorithm \textit{Alg}; time functions $\Thour(\cdot), \Tday(\cdot)$.}
    \For{$t = 0, 1, \ldots$}{
        \vspace{5pt}
        \textbf{\textit{Training phase}} \\
        \For{Bus $n = 1, \ldots, N$}{
            Train the aggregator for $\Ttrain$ steps using \textit{Alg} with initial storage $e^n_t$ under $\vbelief^n_t$ to get $\pi^{n*}_t$ \;
        }
        \vspace{5pt}
        \textbf{\textit{Execution phase}} \\
        \For{Bus $n = 1, \ldots, N$}{
            \ForEach{Prosumer $i = 1, \ldots, M^p_n$}{
                Get net demand $\nd^n_{it} \sim \mathcal{Q}^p_n$ \;
                Take actions $a^n_{it} \sim \pi^{n*}_t$ \;
                Storage state transition as in \eqref{eqn:random-regen} \;
            }
            \For{Consumer $j = 1, \ldots, M^c_n$}{
                Get demand $\nd^n_{jt} \sim \mathcal{Q}^c_n$ \;
            }
            Compute next storage state $e^n_{t+1}$ as in \eqref{eqn:agg-soc-transition} \; 
        }
        Solve an ED problem to get LMPs $\lambda^n_t$ for all $n$ \;
        Update the LMP belief as in \eqref{eqn:lmp-belief} \;
    }
\end{algorithm}

\section{Decentralized Heuristic Algorithm (DHA)}
\label{app:dha}
To serve as a benchmark to compare with the proposed MFE learning framework, we propose a decentralized heuristic algorithm (DHA). At each timestep $t$, each aggregator $n$ determines a storage action by first using the same LMP belief vector as in~\eqref{eqn:lmp-belief}. Then each aggregator $n$ defines the low- and high-price windows using pre-determined and fixed thresholds $\lambda_{\mathrm{low}}^n$ and $\lambda_{\mathrm{high}}^n$ (where $\lambda_{\mathrm{high}}^n>\lambda_{\mathrm{low}}^n$) over the next day:
\begin{align}
\mathcal{T}^n_{\mathrm{low}}(t) := \left\{ h: \left[\vbelief^n_t\right]_h \le \lambda_{\mathrm{low}}^n \right\}, \label{eq:ha_low_window}\\
\mathcal{T}^n_{\mathrm{high}}(t) := \left\{ h: \left[\vbelief^n_t\right]_h \ge \lambda_{\mathrm{high}}^n \right\} \label{eq:ha_high_window},
\end{align}
where $\left[\vbelief^n_t\right]_h$ indicates the $h$-th entry of the vector $\vbelief^n_t$. The cardinality of the two sets can be computed as $N^n_{\mathrm{low}}(t):=|\mathcal{T}^n_{\mathrm{low}}(t)|$ and $N^n_{\mathrm{high}}(t):=|\mathcal{T}^n_{\mathrm{high}}(t)|$. DHA spreads charging across the low-price window rather than charging aggressively at a single hour. If $N^n_{\mathrm{low}}(t)>0$, define the target battery level at the end of the low-price window as
\begin{equation}
e^{n*}_{\mathrm{ch}, t} := \min \left\{1, e^n_t + N^n_{\mathrm{low}}(t)(1-e^n_t)\right\}, \label{eq:ha_target_charge}
\end{equation}
and define the planned per-step charging amount as
\begin{equation}
\overline{a}^C_{nt} := \frac{e^{n*}_{\mathrm{ch}, t}-e^n_t}{N^n_{\mathrm{low}}(t)}. \label{eq:ha_plan_charge}
\end{equation}
Similarly, if $N^n_{\mathrm{high}}(t)>0$, define the target battery level at the end of the high-price window, and the planned per-step discharging amount as follows:
\begin{equation}
e^{n*}_{\mathrm{dis},t} := \max\{0,\ e^n_t - N^n_{\mathrm{high}}(t)\,e^n_t\},
\label{eq:ha_target_dis}
\end{equation}
\begin{equation}
\overline{a}^{D}_{nt} := \frac{e^n_t-e^{n*}_{\mathrm{dis},t}}{N^n_{\mathrm{high}}(t)}. \label{eq:ha_plan_dis}
\end{equation}
Now, given the current level $e^n_t$, the deterministic action $a^{n*}_t\in[-1,1]$ is
\begin{equation}
a^{n*}_t=
\begin{cases}
-\min\{\overline{a}^{D}_{nt},\, e^n_t\}, &\text{if } T_{\mathrm{hour}}(t) \in \mathcal{T}^n_{\mathrm{high}}(t),\\
\min\{\overline{a}^C_{nt},\, 1-e^n_t\}, &\text{if } T_{\mathrm{hour}}(t) \in \mathcal{T}^n_{\mathrm{low}}(t),\\
0, &\text{otherwise},
\end{cases}
\label{eq:ha_det_action}
\end{equation}
where we prioritize discharging if both windows are active. To reduce synchronized behavior, each aggregator draws from an i.i.d. random factor $\xi^n_t \sim \mathrm{Uniform}(\alpha,1)$, where $\alpha\in(0,1)$, and implements the final action $a^n_t := \xi^n_t\, a^{n*}_t$, where in our numerical setup, we choose $\alpha = 0.8$. Finally, the battery level is updated via \eqref{eqn:storage-transition} and the resulting aggregate net demand bids are cleared by the standard ED \eqref{eqn:ed-obj}--\eqref{eqn:gen-limit}. The algorithm pseudo-code is presented in Algorithm~\ref{alg:dha}. 

\begin{algorithm}
    \caption{A Decentralized Heuristic Algorithm}
    \label{alg:dha}
    \KwIn{Initial battery states $e^n_0 \in [0, 1]$, initial LMP beliefs $\vbelief^n_0 \in \mathbb{R}^H$ with learning rates $\delta_n \in [0.5, 1]$, demand shapes $\mathcal{Q}^p_n, \mathcal{Q}^c_n$; thresholds $\{\lambda_{\mathrm{low}}^n,\lambda_{\mathrm{high}}^n\}_{n\in\mathcal{N}}$; randomization parameter $\alpha\in(0,1)$, time functions $\Thour(\cdot), \Tday(\cdot)$.}
    \For{$t = 0, 1, \ldots$}{
        \vspace{5pt}
        \textbf{\textit{Training phase}} \\
        \For{Bus $n = 1, \ldots, N$}{
            Compute $\mathcal{T}^n_{\mathrm{low}}(t), \mathcal{T}^n_{\mathrm{high}}(t)$ via~\eqref{eq:ha_low_window}-~\eqref{eq:ha_high_window}\;
            Compute $\overline{a}^C_{nt}$ and $\overline{a}^{D}_{nt}$ via \eqref{eq:ha_plan_charge}--\eqref{eq:ha_plan_dis}\;
            Set $a^{n*}_t$ via \eqref{eq:ha_det_action} \;
            Draw $\xi^n_t\sim\mathrm{Uniform}(\alpha,1)$ and set $a^n_t=\xi^n_t a^{n*}_t$.
        }
        Submit the implied net demand bids and solve an ED problem to get LMPs $\lambda^n_t$ for all $n$ \;
        Update the LMP belief as in \eqref{eqn:lmp-belief} \;
        Update $\{e^n_{t+1}\}$ using \eqref{eqn:storage-transition}.
    }
\end{algorithm}

\bibliographystyle{ieeetr}
\bibliography{references}
\balance

\endgroup
\end{document}